\documentclass[aip,pra,twocolumn,showpacs]{revtex4}
\usepackage{graphicx,color}
\usepackage{amsmath}
\usepackage{longtable}
\begin{document}

\title{Electron and ion thermal forces in complex (dusty) plasmas}

\author{Sergey A. Khrapak\footnote{
Also at Joint Institute for High Temperatures, Moscow, Russia}
}
\date{\today}
\affiliation{Max-Planck-Institut f\"ur extraterrestrische Physik, D-85741 Garching,
Germany}

\begin{abstract}
Expressions for the ion and electron thermal forces acting on a charged grain, suspended in a weakly ionized plasma subject to temperature gradients, are derived. The main emphasize is on the conditions pertinent to the investigations of complex (dusty) plasmas in gas discharges. Estimates show that for the electron temperature gradients $\sim {\mathcal O}$(eV/cm) typically encountered in laboratory gas discharges, the electron thermal force can become an important player among other forces acting on micron-size grains.
\end{abstract}

\maketitle

\section{Introduction}

The momentum exchange between different species plays an exceptionally important role in complex (dusty) plasmas -- multi-component systems consisting of charged micron-sized (``dust'') grains embedded in a plasma~\cite{Book,FortovPR,MT}. In a weakly ionized plasma, the momentum transfer in collisions with the neutral gas ``cool down'' the system, in particular grains and ions, and introduce some damping. Relative motion between the grains and other plasma components results in the so-called drag forces. The neutral, ion, and (to a lesser extent) electron drag forces are known to affect and often dominate the grain component statical and dynamical properties~\cite{Book,FortovPR,MT,CPP,ED}.

Relative motion between the grain and other plasma species is not the only mechanism which can be responsible for the momentum transfer. A relevant example is the thermophoresis, which
describes the phenomenon wherein small particles, suspended in a gas where a temperature gradient exists (but macroscopic flows are absent), experience a force in the direction opposite to that of the gradient. Elementary consideration of this phenomenon has been given by Einstein~\cite{Einstein} and then it was investigated in detail (see e.g. Refs.~\cite{Waldmann,Talbot,Li} and references therein). In the context of complex plasmas, Jellum {\it et al}.~\cite{Jellum} were apparently the first who recognized the possibility to manipulate the particles in gas discharges using the thermophoretic force. Since then, applying the vertical temperature gradients to compensate for the particle gravity has become a standard technique for controlled particle manipulation in laboratory experiments~\cite{Rothermel,Schwabe2007,Schwabe2009,Schmidt}.

In complex plasmas, the thermophoretic force has its counterparts associated with the charged electron and ion components, provided the corresponding temperature gradients are present. These electron and ion {\it thermal} forces appearing in plasmas have received much less attention than the conventional thermophoresis, or the corresponding drag forces. Brook~\cite{Brook} calculated the force acting on a small {\it uncharged} particle suspended in a plasma in which temperature gradients and a magnetic field are present. Gnedovets~\cite{Gnedovets} took into account the particle charge and demonstrated that the electron and ion thermal forces can have significant effect on the particles which are smaller than the plasma screening length. Nevertheless, these forces have not yet been discussed in the context of complex (dusty) plasmas in any reasonable detail. It seems of value, therefore, to revisit this issue with proper account of the progresses in the understanding of basic plasma-dust interactions achieved in the last two decades.

In this paper the expressions for the ion and electron thermal forces acting on a highly charged grain immersed in a weakly ionized plasma subject to temperature gradients are derived. The focus is on spherical grains, which are smaller than the plasma screening length. Low collisionality is assumed, so that the mean free paths of all plasma species (neutrals, ions, and electrons) are long compared to the grain size, which corresponds to the limit of large Knudsen number. These conditions are typical for complex plasma experiments in laboratory gas discharges. The obtained in this way expressions are simple enough and can be easily implemented for practical applications. The directions and the magnitudes of the forces are analyzed and an interesting and non-trivial behavior is reported. Comparison with other forces indicates that the electron thermal force, in particular, can provide important contributions to the net force balance in typical complex plasma experiments.

\section{Formulation}

The general expression for the force associated with the momentum transfer from a light species $\alpha$ to the massive grain at rest is
\begin{equation}\label{mt}
{\bf F}_{\alpha}=m_{\alpha}\int{\bf v}v\sigma_{\alpha}(v)f_{\alpha}({\bf v})d^3v,
\end{equation}
where $m_{\alpha}$, $f_{\alpha}({\bf v})$ and $\sigma_{\alpha}(v)$ are the corresponding mass, velocity distribution function, and (velocity dependent) momentum transfer cross section ($\alpha = n, i, e$ for neutrals, ions, and electrons, respectively). The net momentum transfer occurs when the velocity distribution has some asymmetry (e.g. relative motion). Assuming weak asymmetry, we write
$f_{\alpha} ({\bf v}) \simeq f_{\alpha 0}(v)+f_{\alpha 1}({\bf v})$, where the symmetric component $f_{\alpha 0}$ is taken to be Maxwellian
\begin{displaymath}
f_{\alpha 0} = n_{\alpha}(m_{\alpha}/2\pi T_{\alpha})^{3/2}\exp(-m_{\alpha}v^2/2T_{\alpha}).
\end{displaymath}
Here $n_{\alpha}$ and $T_{\alpha}$ are the density and temperature (in energy units) of the species $\alpha$. The small asymmetric component $f_{\alpha 1}$, which gives contribution to the integral in (\ref{mt}), depends on the nature of the anisotropy. In the case of subthermal drifts with relative velocity ${\bf u}_{\alpha}$ ($u_{\alpha}\lesssim v_{T_{\alpha}}$) it reduces to $f_{\alpha 1}\simeq f_{\alpha 0}({\bf v}{\bf u}_{\alpha}/v_{T_{\alpha}}^2)$, where $v_{T_{\alpha}}=\sqrt{T_{\alpha}/m_{\alpha}}$ is the thermal velocity. This ansatz corresponds to the conventional calculation of the neutral, ion and electron drag forces for subthermal flows and has been thoroughly investigated earlier~\cite{FortovPR,MT,ED,Epstein,id2002}.

The focus of the present paper is on a complementary situation when relative drifts are absent, but the momentum transfer do occur due to the net momentum flux caused by the temperature gradients (If both temperature gradients and subthermal drifts are present, the corresponding thermal and drag forces are added in a simple superposition.) In this case, the asymmetric part of the velocity distribution function of the component $\alpha$ can be approximated as
\begin{equation}\label{f1}
f_{\alpha 1}\simeq \frac{m_{\alpha}\kappa_{\alpha}f_{\alpha 0}}{n_{\alpha}T_{\alpha}^2}\left[1-\frac{m_{\alpha}v^2}{5 T_{\alpha}}\right]{\bf v}\nabla T_{\alpha},
\end{equation}
where $\kappa_{\alpha}$ is the thermal conductivity of the species $\alpha$. This expression can be for instance derived by linearizing the kinetic equation with the BGK-like collision operator and expressing the effective collision frequency via the thermal conductivity. It is easy to check that this form ensures that the self-consistent density gradient and/or electric field (for charged species), build up in response to the temperature gradient, result in no net flux: ${\bf j}_{\alpha}=\int{\bf v}f_{\alpha 1}d^3v = 0$ (i.e., ${\bf u}_{\alpha}=0$). The Fourier's law for heat transfer is also satisfied: ${\bf q}_{\alpha}=\int(m_{\alpha}v^2/2){\bf v}f_{\alpha 1}d^3v = -\kappa_{\alpha}\nabla T_{\alpha}$. Equation (\ref{f1}) is an approximation, which is exact only for the special case of $\propto r^{-4}$ interactions (for the rigorous mathematical treatment of non-uniform gases see Ref.~\cite{Chapman}). Its accuracy is, however, more than acceptable for our present purposes, especially in view of further simplifications involved in treating electron-grain and ion-grain collisions.

Substituting Eq.~(\ref{f1}) into (\ref{mt}) we get after integration over the angles in spherical coordinates
\begin{equation}\label{force}
F_{{\rm T}\alpha}=\frac{16\kappa_{\alpha}\nabla T_{\alpha}}{15\sqrt{2\pi}v_{T_{\alpha}}}\int_0^{\infty}x^2(\tfrac{5}{2}-x)\exp(-x)\sigma_{\alpha}(x)dx,
\end{equation}
where the remaining integration is over the reduced velocity $x=v^2/2v_{T_{\alpha}}^2$. For grain-neutral collisions the momentum transfer cross section is velocity-independent, $\sigma_n(x)=\pi a ^2$, and the integration yields $F_{{\rm T}n}= - (8\sqrt{2\pi}/15)(\kappa_n a^2/v_{T_n})\nabla T_n$, where $a$ is the particle radius. This coincides with the celebrated expression by Waldmann~\cite{Waldmann} for the conventional thermophoretic force. The force is directed towards lower gas temperature, because hotter atoms transfer more momentum to the grain than the colder ones.

It is convenient to write the generic expression for the thermal forces in complex plasmas as
\begin{equation}\label{generic}
F_{{\rm T}\alpha}=-\frac{8\sqrt{2\pi}}{15}\frac{\kappa_{\alpha}a^2\nabla T_{\alpha}}{v_{T_{\alpha}}}\Phi_{\alpha}.
\end{equation}
For the neutral component (thermophoresis) $\Phi_n=1$. The factors $\Phi_{i(e)}$ account for the electrical interactions between the ions (electrons) and the charged grain. They depend on the shape of the electrical potential around the grain. We use the conventional Debye-H\"{u}ckel (Yukawa) form, $\phi(r) \simeq (Q/r)\exp(-r/\lambda)$, where $Q$ is the grain charge and $\lambda$ is the plasma screening length. In low-temperature gas discharges the dominant charging process is the continuous absorption of electrons and ions on the grain surface. In this case the charge is negative and proportional to the product of the grain radius and the electron temperature, viz. $Q = -z(aT_e/e)$, where $z$ is the coefficient of order unity, which depends on the plasma parameter regime~\cite{CPP}. In the absence of substantial ion flows and strong nonlinearities in ion-grain interaction, the screening length is approximately given by the ion Debye radius $\lambda\simeq \lambda_{{\rm D}_i}=\sqrt{T_i/4\pi e^2 n_i}$, provided $T_e\gg T_i$. When ion-grain interaction is strongly non-linear, the effective screening length exceeds the ion Debye radius~\cite{Daugherty}, approximate fits are available in the literature~\cite{CPP}. Although it is well recognized that the long range asymptote of the electrical potential can be modified (by e.g. continuous plasma absorption on the grain surface~\cite{TsytovichUFN,Allen2000,FilippovJETPL2007,KKM} or plasma ionization/recombination effects~\cite{IR}), this is expected to affect merely grain-grain interactions, but not the momentum transfer from the ions and electrons.

\section{Electron component}

Normally, the grain radius in complex plasmas is much smaller than the plasma screening length. In this case, the electron-grain interaction can be called ``weak'', in the sense that its range -- the Coulomb radius $R_{\rm C}^e\sim za$  -- is much smaller than the screening length $\lambda$~\cite{MT}. This implies that the Coulomb scattering theory is appropriate to describe electron-grain elastic collisions. The general momentum transfer cross section for scattering in the Coulomb potential $Q/r$ is
\begin{equation}\label{sigma}
\sigma_{\alpha}^{\rm s}(v)=4\pi R_{\alpha}^2(v)\ln\left[\frac{R_{\alpha}^2(v)+\rho_{\rm max}^2(v)}{R_{\alpha}^2(v)+\rho_{\rm min}^2(v)}\right]^{1/2},
\end{equation}
where $R_{\alpha}(v)= (|Q|e/m_{\alpha}v^2)$ and $\alpha=e,i$. The maximum impact parameter $\rho_{\rm max}$ is necessary to avoid the logarithmic divergence of the cross section. In the standard Coulomb scattering theory $\rho_{\rm max}=\lambda$, and this choice is appropriate for the weak electron-grain interaction~\cite{MT,ED}. The minimum impact parameter $\rho_{\rm min}$ is zero in the standard Coulomb scattering theory. In the considered case, however, some electrons are falling onto the grain instead of being elastically scattered in its electrical potential. Consequently, $\rho_{\rm min}$   should be set equal to the maximum impact parameter corresponding to the electron collection by the grain, $\rho_{\rm c}$. The orbital motion limited (OML) theory yields~\cite{Allen} $\rho_{\rm c}(v)=a\sqrt{1-2R_e(v)/a}$ for $2R_e(v)<a$ and $\rho_{\rm c}=0$ otherwise (sufficiently slow electrons cannot be collected even in head-on collisions with the grain due to electrical repulsion). The momentum transfer cross section for electron collection is simply $\sigma_e^{\rm c}(v)=\pi \rho_{\rm c}^2(v)$. Expressing now velocity in terms of $x=v^2/2v_{T_e}^2$, substituting $\sigma_e(x)=\sigma_e^{\rm s}(x)+\sigma_{e}^{\rm c}(x)$ into Eq.~(\ref{force}) and performing the integration with the appropriate limits yields:
\begin{equation}\label{etf}
\Phi_e=(1+\tfrac{3}{2}z+z^2)\exp(-z)-\tfrac{z^2}{2}\Lambda_e,
\end{equation}
where $\Lambda_e$ is the electron-grain Coulomb logarithm
\begin{equation}\label{ECL}
\Lambda_e=\int_0^{\infty}h(x)\ln\left(1+\tfrac{4\lambda^2}{a^2}\tfrac{x^2}{z^2}\right)dx \\-2\int_z^{\infty}h(x)\ln\left(\tfrac{2x}{z}-1\right)dx.
\end{equation}
The function $h(x)$ is defined as $h(x)=\left(\tfrac{5}{2}-x\right)\exp(-x)$. Equations (\ref{generic}), (\ref{etf}) and (\ref{ECL}) constitute the expression for the electron thermal force. It is similar to that obtained earlier by Gnedovets~\cite{Gnedovets}. To get identical expressions one needs to neglect unity compared to $\lambda/a$ in the equations of Ref.~\cite{Gnedovets}.

\begin{figure}
\includegraphics[width=6.5cm]{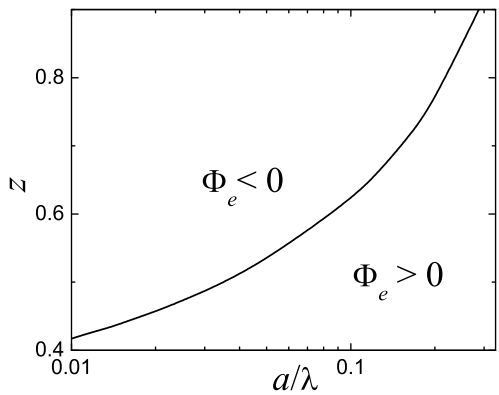}
\caption{The transition curve separating regions of positive and negative electron thermal force in the plane of reduced grain size $a/\lambda$ and charge $z$.}
\label{Fig1}
\end{figure}

For an uncharged particle $\Phi_e=1$ and the electron thermal force pushes the grains in the direction of lower electron temperature, similarly to the thermophoretic force. According to Eq.~(\ref{etf}) the contributions from collection and scattering are directed oppositely to each other. For a sufficiently high charge, the scattering part becomes dominant and the force reverses direction. The physical reason is that the (Coulomb) scattering momentum transfer cross section quickly decreases with velocity [$\propto v^{-4}$, see Eq. (\ref{sigma})], so that the cold electrons are more effective in transferring their momentum upon scattering. In this regime ($\Phi_e<0$), the thermal force acts in the direction of higher electron temperature. In Figure~\ref{Fig1} the curve separating the positive and negative values of $\Phi_e$ is plotted in the plane ($a/\lambda$, $z$). For most experimental conditions $\Phi_e<0$ should be expected, although the transition line does not seem unreachable~\cite{Comment1}.

\section{Ion component}

The Coulomb radius of ion-grain interaction $R_{\rm C}^i\sim za\tau$ is not necessary small compared to the plasma screening length due to the presence of a (normally) large factor $\tau=T_e/T_i$ -- electron-to-ion temperature ratio~\cite{MT}. The interaction range can exceed $\lambda$ and considerable amount of momentum transfer can occur for impact parameters beyond $\lambda$. This implies that the standard Coulomb scattering approach is inappropriate~\cite{id2002,id2003}. It makes sense to consider two regimes of ion scattering separately.
In the regime of {\it moderate ion-grain interaction}, an extension of the standard Coulomb scattering theory is possible by taking into account the momentum transfer from the ions that {\it approach} the grain closer than $\lambda$~\cite{id2002}. This results in $\rho_{\rm max}=\lambda\sqrt{1+2R_i(v)/\lambda}$. This approximation demonstrates good accuracy for $R_i(v)\lesssim 5\lambda$ and reduces to $\rho_{\rm max}=\lambda$ in the limit of weak ion-grain interaction [$R_i(v)\ll \lambda$]. The impact parameter corresponding to the ion collection is $\rho_{\rm c}(v)=a\sqrt{1+2R_i(v)/a}$ and the corresponding collection cross section is $\sigma_{\rm c}(v)=\pi\rho_{\rm c}^2(v)$. Combining the contributions from collection and scattering yields:
\begin{equation}
\Phi_i = 1-\tfrac{1}{2}z\tau-z^2\tau^2\Lambda_i,
\end{equation}
where $\Lambda_i$ is the (modified) ion-grain Coulomb logarithm
\begin{equation}
\Lambda_i=\int_0^{\infty}h(x)\ln\left[\frac{2x(\lambda/a)+z\tau}{2x+z\tau}\right]dx.
\end{equation}
In typical complex plasmas with $\lambda\gg a$, $\tau\gg 1$, and $z\sim 1$, the Coulomb logarithm can be roughly estimated as $\Lambda_i\simeq \ln(1+\beta_T^{-1})$, where $\beta_{T}= \beta(v_{T_i})= (a/\lambda)z\tau$ and $\beta(v)=R_i(v)/\lambda$ is the ion scattering parameter~\cite{id2002}. The approach is reliable up to $\beta_{T}\lesssim 5$. Since the product $z\tau$ is normally quite large, $z\tau \sim {\mathcal O}(10^2)$, it follows that (i) the scattering provides dominant contribution to the momentum transfer; (ii) $\Phi_i<0$, i.e. the grains are pushed into the region with higher ion temperature. The physical reason is again fast decrease of the scattering momentum transfer cross section with the ion velocity, so that cold ions transfer more momentum to the grain. This is not a unique example when the force acting on a charged grain is directed oppositely to the net ion momentum flux. Another example is related to the sign reversal of the ion drag force acting on an absorbing particle in the highly collisional (continuum) limit~\cite{CPP,KhrapakJPD,VladimirovPRL,Filippov}, although the detailed physics is different.

In the regime of very {\it strong ion-grain interaction}, the scattering is characterized by the formation of a potential barrier for ions with impact parameters above the critical one $\rho_*$, which considerably exceeds $\lambda$. For $\rho<\rho_*$ scattering with large angles occurs, which gives the major contribution to the momentum transfer. Relative importance of momentum transfer from distant collisions with $\rho>\rho_*$ decreases rapidly with the increase in ion-grain interaction strength. The detailed consideration of the momentum transfer in this regime can be found in Refs.~\cite{id2003,idIEEE}. In the limit $R_i(v)\gg \lambda$ (i.e. $\beta(v)\gg 1$~\cite{Comment3}) the total momentum transfer cross section (collection and scattering) can be roughly approximated as $\sigma_{\Sigma}(v)\simeq \pi\rho_*^2(v)$, with $\rho_*^2\simeq \lambda^2\left\{\ln^2[\beta(v)]+2\ln[\beta(v)]\right\}$. The integration yields
\begin{equation}
\Phi_i\simeq -\left(\tfrac{\lambda}{a}\right)^2\int_0^{x_*}x^2h(x)\left[\ln^2\left(\tfrac{\beta_T}{2x}\right)+2\ln\left(\tfrac{\beta_T}{2x}\right)\right]dx.
\end{equation}
The upper limit of integration can be chosen as $x_*=\beta_T/2$ to avoid unphysical regime of negative cross section in this approximation. However, due to the presence of the exponentially small term in $h(x)$, taking $x_*=\infty$ will not produce big errors for $\beta_T\gg 1$. Note that the sign of $\Phi_i$ is changed from negative to positive upon increasing $\beta_T$ (in the considered approximation for $\sigma_{\Sigma}(v)$ this happens at $\beta_T\simeq 120$; the exact value is rather sensitive to the functional dependence of the cross section on the ion velocity and, thus, is subject to significant uncertainty). Physically, the sign reversal occurs because scattering in the Yukawa potential in the limit of strong interaction tends to that on a hard sphere with the radius $\rho_*$, which only weakly (logarithmically) depends on the ion velocity.

\section{Discussion}

Having derived the expressions for the ion and electron thermal forces, let us discuss their relative importance. The relationship between $F_{{\rm T}e}$ and $F_{{\rm T}i}$ depends on many factors and is, in principle, arbitrary. In weakly ionized plasmas $\kappa_{\alpha}\sim (n_{\alpha}/n_n)(v_{T_{\alpha}}/\sigma_{{\alpha}n})$  (where $\sigma_{\alpha n}$ is the transport cross sections for collisions with neutrals) and the force ratio becomes $|F_{{\rm T}i}/F_{{\rm T}e}|\sim (\sigma_{en}/\sigma_{in})(\Phi_i/\Phi_e)(\nabla T_i/\nabla T_e)$, where the quasineutrality condition $n_i\sim n_e$ has been used. Assuming further that most of the contribution to the thermal forces comes from elastic scattering, and setting $\Lambda_i\sim \Lambda_e\sim 1$ for simplicity, we end up with
$|F_{{\rm T}i}/F_{{\rm T}e}|\sim (\sigma_{en}/\sigma_{in})\tau^2$ for comparable temperature gradients. The first factor is small, whilst the second can be quite large, and thus various situations are possible.

Another situation deserving attention is when a gradient of the neutral gas temperature is created (e.g. by heating the lower electrode of a parallel plate rf discharge) to compensate for grain gravity (thermophoretic levitation)~\cite{Jellum,Rothermel}. In this case, the ion temperature is likely coupled to the neutral gas temperature, $\nabla T_i\simeq \nabla T_n$. The neutral and ion thermal forces are directed in the opposite directions (in the regime of weak and moderate ion-grain coupling). Their ratio is $|F_{{\rm T}i}/F_{{\rm T}n}|\sim (n_i/n_n)(\sigma_{nn}/\sigma_{in})z^2\tau^2\Lambda_i$. Assuming $\sigma_{in}\sim \sigma_{nn}$~\cite{Comment2} and $\Lambda_i\sim 1$ we find that $F_{{\rm T}i}$ dominates over $F_{{\rm T}n}$ for $n_i/n_n\gtrsim (z\tau)^{-2}\sim 10^{-4}$. For a more typical (in laboratory gas discharges) ionization fraction $n_i/n_n\sim 10^{-6}$, the ion thermal force is only slightly reducing the effect of thermophoretic force.

It is worth to remind that in laboratory gas discharges, ions produced in the plasma bulk are drifting towards the walls and electrodes of the chamber. So even if some gradients of the ion temperature are present, the related effects are likely masked by the more pronounced ion drag force associated with these drifts. For example, consider the expression for the ion drag force derived in Ref.~\cite{id2002} for the subthermal ion drift regime and weak-to-moderate ion-grain coupling, $F_{\rm id}\simeq (8\sqrt{2\pi}/12)a^2n_im_iv_{T_i}u_iz^2\tau^2\Lambda$ ($\Lambda$ is the Coulomb logarithm for ion scattering, which gives the dominant contribution to the ion drag force), along with the relation between the ion drift velocity and the electric field $u_i\simeq (eE/m_i\nu_{in})$, where $\nu_{in}\simeq n_nv_{T_i}\sigma_{in}$ is the characteristic momentum transfer frequency. The Coulomb logarithms encountered in the calculations of the ion drag and ion thermal forces are not exactly the same, but are comparable $\Lambda\sim\Lambda_i$. This results in a particularly simple estimate $|F_{\rm id}/F_{{\rm T}i}|\sim eE/\nabla T_i$. The electric fields of the order of 1 V/cm are quite natural even for the bulk regions of gas discharges (ambipolar fields). The ion temperature is normally close to the neutral gas (room) temperature, and thus for most situations $\nabla T_i$ is orders of magnitude less than 1 eV/cm. This implies that in the considered case  $|F_{\rm id}/F_{{\rm T}i}|\gg 1$.

Electrons are also drifting towards the discharge walls and electrodes. In the regime of ambipolar diffusion electron and ion fluxes are equal to each other. This implies that the electron drag force is negligible as compared to the ion drag force in this regime, due to small electron mass. However, electron temperature gradients can result in the thermal force acting on the particles. The existence of significant gradients of $T_e$  is known from earlier spatially resolved probe measurements of the electron energy distribution function in various gas discharges~\cite{Godyak,Fusselman}. For devices used in complex plasma research, the more recent results from probe measurements~\cite{Wolter}, optical emission spectroscopy~\cite{Mitic}, as well as from numerical modeling~\cite{LandNJP2007,Arp} all revealed gradients in $T_e$ of the order of ${\mathcal O}$(eV/cm). The spatial distribution of $T_e$ generally depends on discharge geometry, plasma parameters, and can be affected by the presence of grains~\cite{LandNJP2007}. Let us therefore assume $\nabla T_e=1$ eV/cm and restrict ourselves to the comparison of the absolute magnitudes of the electron thermal force and other forces acting on a small grain in the bulk of a gas discharge. We take the plasma parameter set from Ref.~\cite{PoP2005} used to estimate relative importance of the electrical and ion drag forces: argon gas at a pressure $p=10$ Pa ($n_n\simeq 2\times 10^{15}$ cm$^{-3}$), $n_e=n_i=3\times 10^9$ cm$^{-3}$, $T_e= 1$ eV, $T_i=T_n=0.03$ eV,  $a=1$ $\mu$m ($a/\lambda\simeq 0.04$), and $z\simeq 3$ (estimated from the collisionless OML theory).  From Eqs. (\ref{etf}) and (\ref{ECL}) we get $\Phi_e\simeq -22$. The electron thermal conductivity in a weakly ionized plasma is $\kappa_e\simeq \tfrac{5}{2}(n_e T_e/m_e\nu_{en})$ with $\nu_{en}\simeq n_n\sigma_{en}v_{T_e}$ ($\sigma_{en}\sim 10^{-16}$ cm$^2$ for $T_e\sim 1$ eV in argon). This results in $F_{{\rm T}e}\simeq 2\times 10^{-8}$ dyne. This force is more than three times larger than the force of gravity, experienced by the grain of this size (and material density of $1.5$ g/cm$^3$) in ground-based experiments. The equivalent electric field $E_*$, for which $F_{{\rm T}e}\simeq |Q|E_*$, is $E_*\simeq 5$ V/cm. Such a field would produce a significant plasma anisotropy, characterized by superthermal ion flows (Fig. 4b from Ref.~\cite{PoP2005}). Finally, this magnitude is comparable to the {\it maximum} value of the ion drag force the grain can experience in subsonic ion flows for this set of parameters ($\sim 5 \times 10^{-8}$ dyne according to Fig. 4a from Ref.~\cite{PoP2005}).

Observation of big grains, trapped in standing striations of a stratified dc glow discharge~\cite{Lipaev}, provides another example where the electron thermal force can play a significant role. The electron temperature is known to increase considerably in the head of the striation (where their energy is of the order of the first excitation potential), which can explain that even very massive grains can be confined there~\cite{Lipaev}.

Other situations wherein plasma thermal forces may play significant role include charged aerosols, dust in planetary (e.g., Earth) atmospheres, fusion devices, and even quark-gluon plasma~\cite{Thoma}. Present results may not, however, be directly applicable to some of these cases. For example, the regime of fully ionized magnetized plasma, considered recently~\cite{Stepanenko}, is apparently more relevant in the context of dust in fusion devices. As a final remark, we point out that complex plasmas represents a natural example where ''negative thermophoresis'' (force pointing towards higher temperatures) can exists. A theoretical criterion for negative thermophoresis in dilute non-ionized gases has been discussed in a recent publication~\cite{Wang}.

\section{Conclusion}

To conclude, the forces acting on small particles in a weakly ionized plasma subject to ion and electron temperature gradients have been analyzed with the main emphasis on applications to complex (dusty) plasmas. The presented estimates demonstrate that the ion thermal force is usually of minor importance in gas discharges presently employed in complex plasma research. In contrast, the electron thermal force can well be comparable to other forces acting on micron-size particles. This finding should be properly addressed when developing new (and updating existing) numerical codes to model particle transport in gas discharges. The expressions derived in this paper are easy for practical implementation and can serve as a theoretical basis for further detailed studies.


\begin{thebibliography}{99}
\bibitem{Comment2} For ions in their parent gases $\sigma_{in}$ is normally several times larger than $\sigma_{nn}$ due to resonant charge exchange.
\end{thebibliography}
\end{document}